\documentclass[12pt]{article}
\usepackage{epsfig}

\def\beq{\begin{equation}}
\def\eeq#1{\label{#1}\end{equation}}
\def\eeqn{\end{equation}}

\def\beqa{\begin{eqnarray}}
\def\eeqa#1{\label{#1}\end{eqnarray}}
\def\eeqan{\end{eqnarray}}

\let\bar=\overbar

\def\etal{{\it et al.}}

\def\bra#1{\left\langle{ #1} \right|}
\def\ket#1{\left| {#1} \right\rangle}

\def\half{\frac{1}{2}}

\def\Dslash{\not{\hbox{\kern-4pt $D$}}}
\def\dslash{\not{\hbox{\kern-2pt $\del$}}}

\def\ee{e^+e^-}

\def\msb{{\bar{\ssstyle M \kern -1pt S}}}

\def\vus{V_{us}}
\def\vud{V_{ud}}
\def\epe{\epsilon'/\epsilon}

\def\Title#1{\begin{center} {\Large {\bf #1} } \end{center}}

\begin{document}

\Title{Recent results on $\vus$ from KLOE, KTeV and NA48}

\begin{center}{\large \bf Contribution to the proceedings of HQL06,\\
Munich, October 16th-20th 2006}\end{center}

\bigskip\bigskip


\begin{raggedright}  

{\it Venelin Kozhuharov\index{Kozhuharov, V.}\\
Atomic Physics Department, \\
Faculty of Physics, University of Sofia\\
5 J. Bourchier Str., Sofia, BULGARIA
\footnote{Also JINR Dubna, 141980, Dubna, Moscow Region, RUSSIA} }
\bigskip\bigskip
\end{raggedright}

\section{Introduction}

The flavour structure in the quark sector of the Standard Model is described by the CKM matrix \cite{c},\cite{km}. Its unitarity leads to a number of relations for its elements and in particular for the first raw:
\begin{equation}
|V_{ud}|^2 + |\vus |^2 + |V_{ub}|^2 = 1
\end{equation}
Since $V_{ub}\cong 4\times10^{-3}$ the contribution of the last term could be neglected at the current level of uncertainty in $V_{ud}$ and $\vus$. This approximation gives $\vus=\sin \theta_c$ as originally suggested by Cabibbo. 

The most precise value of $V_{ud}$ comes from the super-allowed $0^+ \to 0^+$ beta transitions between nuclei and $\vus$ is usually calculated from the branchings of the kaon semileptonic decays.  
Going back to PDG 2004 \cite{pdg2004} $\vus=0.2195\pm 0.0025$ and $V_{ud} = 0.9738 \pm 0.0005$ giving a deviation from unitarity at the level of $2.3\sigma$ where the contribution from the uncertainties of $V_{ud}$  and $\vus$ in the final error are almost equal.  

In the last few years a significant progress in the kaon physics has been made by three experiments - KLOE, KTeV and NA48. The reflection of their results to the extraction of $\vus$ is subject of this review.

KTeV at the Main Injector (Fermilab) \cite{ktevtechnical}  and NA48 at SPS (CERN) \cite{na48technical} are fixed target experiments and exploit similar techniques of kaon decays in flight. Both consist of a spectrometer system measuring the charged particles momentum and a calorimetry system  used for measurement of the energy of photons and electrons. The calorimetry system also provides a way to distinguish between the different type of charged particles through their interactions with matter. A muon veto system is placed at the end of each detector complex. The primary purpose of both experiments was to measure the direct CP violation parameter $\epe$ in the neutral kaon system \cite{epove}. In 2003 NA48 modified its setup in order to study charged kaon decays. 

KLOE experiment \cite{kloetechnical} is  situated at $DA\Phi NE$, the Frascati $\phi$ factory, where  $\ee$ beams collide with a center of mass energy at the $\phi$ meson mass (1020 MeV). With a probability of $\approx 83\%$  $\phi$ decays into neutral or charged kaons, anticolinear in the $\phi$ center of mass (almost true also in the laboratory system). The presence of $K_{L/S}$  ($K^{\pm}$)   tags $K_{S/L}$ ($K^{\mp}$). KLOE detector has $2\pi$ symmetry, the momentum of the decay products is measured by a magnetic spectrometer which is followed by an elecromagnetic calorimeter. 

\section{Kaon semileptonic decays}
Within the Standard Model $K\to \pi l \nu$  (so called $Kl3$) decay appear as a tree level process of $ s \to u $ transition. The inclusive branching ratios of all four modes ($K^0e3$, $K^0\mu3$, $K^{\pm}e3$ and $K^{\pm}\mu3$) could be written conveniently in the form 
\begin{equation}
Br(K_{l3(\gamma)}) =  \frac{G_F^2 M_K^5 S_{EW}} {128 \pi ^3/ \tau _K  } I_K C^2 (1+ \delta^I_{EM}  ) \times | V_{us} {f^{K\pi}_+ (0) }|^2
\end{equation}
where $G_F$ is the Fermi constant, $M_K$ and $\tau_K$ are the corresponding kaon mass and lifetime,  $S_{EW}$ is the the short distance electroweak enhancement factor, $S_{EW} \cong 1+ \frac{2\alpha}{\pi} (1 - \frac{\alpha_s}{4\pi}) \times log\frac{M_Z}{M_\rho}  =  1.023$ \cite{marsir}, C is the Klebsh-Gordon coefficient , $C=1$ for $K^0$ and $C=\sqrt{\half}$ for $K^{\pm}$ , $\delta^I_{EM}  $ represents the long-distance electromagnetic correction \cite{ciriglianokch,ciriglianok0}, ${f^{K\pi}_+ (0) }$ is the value of the vector form-factor at zero transferred momentum and $I_K$ is the phase space integral dependent on the mode and the shape of the form-factor.  

The values of $S_{EW}$, $\delta^I_{EM}  $ and ${f^{K\pi}_+ (0) }$ are calculated theoretically while the rest could be obtained from experimental measurements. 

\subsection{Form factors}
The kaon form factors are defined as \cite{ffs}
\begin{equation}
\bra{\pi(q)} s\gamma_{\mu} u  \ket{K(p)} = f^{K\pi}_+ (t)\times (p+q) + f^{K\pi}_- (t) \times (p-q)
\end{equation}
where $t=(p-q)^2$ is the transferred momentum. Instead of the couple $f_+, f_-$ usually another set of form-factors is used $f_+(t)$ and $f_0(t) = f_+(t) + \frac{t}{M_K^2 - M_{\pi}^2} f_-(t)$ inspired by the VMD model.
 The dependence of the transferred momentum could be written as
\begin{equation}
f^{K\pi}_{+,0} (t) = f^{K\pi}_+(0) (1 + \delta f_{+,0}(t))
\end{equation}
It is convenient to express the charged kaon form factor by the neutral one $|f^{K^+\pi^0}_+(0)|^2 = (1+\delta_{SU2})\times  |f^{K^0\pi^-}_+(0)|^2$. The SU2 breaking parameter is obtained within the Chiral Perturbation Theory,  $\delta_{SU2}=0.046 \pm 0.004$ \cite{ciriglianokch,leutroos}.
$f_+(0)$ was calculated for the first time in the 80s \cite{leutroos}
\begin{equation}
f_+(0) = 0.961 \pm 0.008.
\label{fplus}
\end{equation}
However more recent analysis give higher values $f_+(0) = (0.981 \pm 0.012)$ \cite{ciriglianok0}. 
Another result  $f_+(0) = (0.960 \pm 0.009)$ comes from lattice QCD \cite{fpluslattice} which is consistent with (\ref{fplus}). Since $f_+(0)$ enters directly in the calculation of $\vus$ a clarification of this problem is highly desirable. In this review (\ref{fplus}) is used.

The term $\delta f_{+,0}(t)$ enters in the phase space integral calculation and is subject to different parametrization. The Taylor expansion  gives
\begin{equation}
\delta f_{+,0}(t) = \lambda_{+,0} ' \frac{t}{M_{\pi}^2} + \half \lambda_{+,0} " \frac{t^2}{M_{\pi}^4}.
\label{ffquad}
\end{equation}
while within the VMD model  $f_{+,0}$ correspond to vector or scalar meson exchange and are parametrized  by the mass of the pole:
\begin{equation}
\delta f_{+,0}(t)  = \frac{M_{V,S}^2}{M_{V,S}^2 - t} -1
\end{equation}
In both cases the unknown parameters are determined experimentally. If in equation (\ref{ffquad}) the quadratic term is neglected then the shape of the form factor is given only by its slope $\lambda_+$. The three collaborations have studied the form factors in the case of $K_L \to \pi^0 e \nu $ decays and the results can be summarized in the following table:

\begin{table}[h!]
\label{formfactors}
\begin{tabular}{|l|c|c|c|c|}
\hline 
\hline
      & $\lambda_+ ' $ & $\lambda_+ "$ & $\lambda_+$ & Pole mass \\ \hline 
NA48 \cite{na48ff} & $0.0280 \pm 0.0024 $ &   $ 0.0004 \pm 0.0009 $  & $0.0288 \pm 0.0012$  &$ 859 \pm 18$  \\ \hline
KTeV \cite{ktevff} & $ 0.0217 \pm 0.0020 $   & $ 0.0029 \pm 0.0008 $ &  $0.0283 \pm 0.0006$ &$ 881 \pm 7.1 $  \\ \hline
KLOE \cite{kloeff} & $0.0255 \pm 0.0018 $  &  $ 0.0014 \pm 0.0008  $ & $0.0286 \pm 0.0006$ &$ 870 \pm 9.2 $  \\ \hline
\hline
\end{tabular}
\end{table}
The values agree in the case of linear and pole parametrization but there is a discrepancy for the necessity of a quadratic term in (\ref{ffquad}). Recently the KTeV collaboration has performed a new calculation of the phase space integral with a reduced model uncertainty, $ I_{K0e3} = 0.10262 \pm 0.00032 $ \cite{ktevps}. For the rest of the phase-space integrals we use $I_{K0\mu3} = 0.06777 \pm 0.00053  $ with the KTeV quadratic form factor parametrization,  $ I_{K^{\pm}e3} = 0.1060 \pm 0.0008 $ and $ I_{K^{\pm}\mu 3} = 0.0702 \pm 0.0005 $ with the ISTRA+ measurement of the form factors \cite{istraff}.  A 0.7\% error is added to account for the difference between the quadratic and the pole parametrization of the form-factors. 

\subsection{Kaon lifetime}
During the last year two new measurements of the $K_L$ lifetime have been published by KLOE. One of them is obtained from the the proper time distribution of $K_L \to 3\pi^0$ decays \cite{kloelife},  giving  $\tau_{KL} = (50.92\pm 0.30) ns$.  The second method produces a result for the lifetime as a byproduct of the measurement of the major $K_L$ branching fraction imposing the condition that their sum should be unity \cite{kloeklbr}. The result is $\tau_{KL} = (50.72 \pm 0.37) ns $, independent of the previous measurement. The combined value including also the only previous measurement in the 70s is  $\tau_{KL} = (51.01 \pm 0.20) ns$. For the $K_S$ lifetime the PDG \cite{pdg06} average is used.

Concerning the charged kaons a new preliminary result for the $K^{\pm}$ lifetime has been presented by KLOE $\tau_{K\pm} = (1.2367\pm 0.0078) \times 10^{-8} s $ \cite{kloehql06}. For the moment the PDG average $ \tau_{K\pm} = (1.2385 \pm 0.0025)  \times 10^{-8} s  $ is used and we are waiting for the final result. 

\subsection{Branching ratios}
For a long time the branching ratios of the kaon semileptonic decays were fixed in the PDG due to the lack of new measurements. The BNL result for $Br(K^+e3) = (5.13\pm0.10)\%$ \cite{bnlke3} published in 2003 was in disagreement with the PDG 2002 value ($Br(K^+e3) = (4.87\pm0.06)\%$ ) \cite{pdg02} and initiated a lot of experimental activity. 

All six major $K_L$ branching fractions have been measured by KTeV determining their ratios of decay rates \cite{ktevbrall}. The results for $Br(K_L e3)$ and  $Br(K_L \mu3)$ are
\begin{eqnarray}
Br(K_L \to \pi^{\pm} e^{\mp} \nu) = (40.67 \pm 0.11) \% \\
Br(K_L \to \pi^{\pm} \mu^{\mp} \nu) = (27.01 \pm 0.09) \% 
\end{eqnarray}

KLOE has also measured the dominant $K_L$ branchings \cite{kloeklbr} as mentioned above obtaining for the semileptonic decays
\begin{eqnarray}
Br(K_L \to \pi^{\pm} e^{\mp} \nu) = (40.07 \pm 0.15) \% \\
Br(K_L \to \pi^{\pm} \mu^{\mp} \nu) = (26.98 \pm 0.15) \% 
\end{eqnarray}
Apart from the $K_L$ KLOE has studied  $K_S e3$ decays \cite{kloeks}. Using $K_S \to \pi^+\pi^-$ for normalization channel the result is four times more precise than the previous value:
\begin{equation}
Br(K_S \to \pi^{\pm} e^{\mp} \nu) = (7.046 \pm 0.091) \% 
\end{equation}

\begin{figure}[!ht]
\includegraphics[width=75mm]{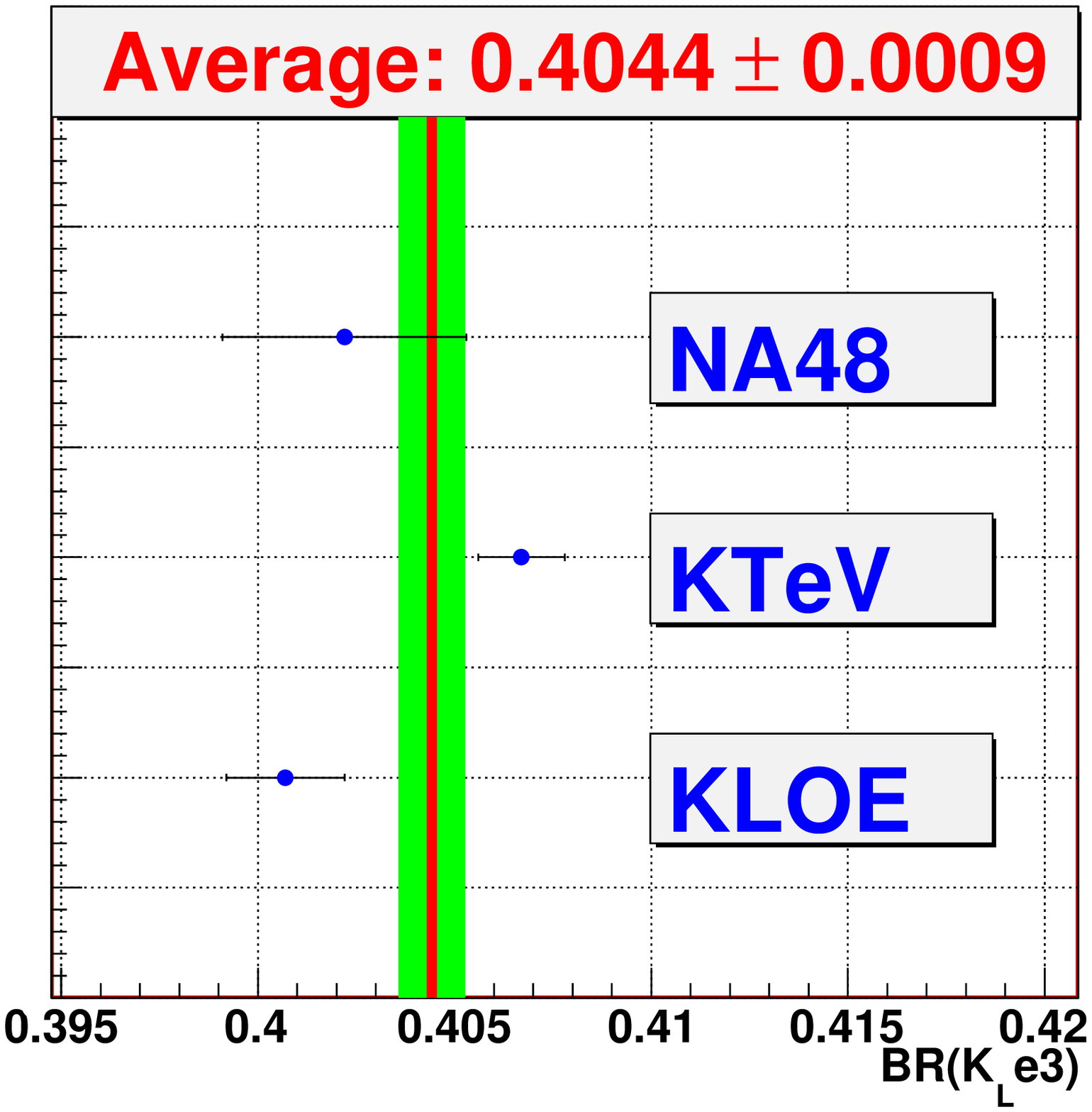}
\includegraphics[width=75mm]{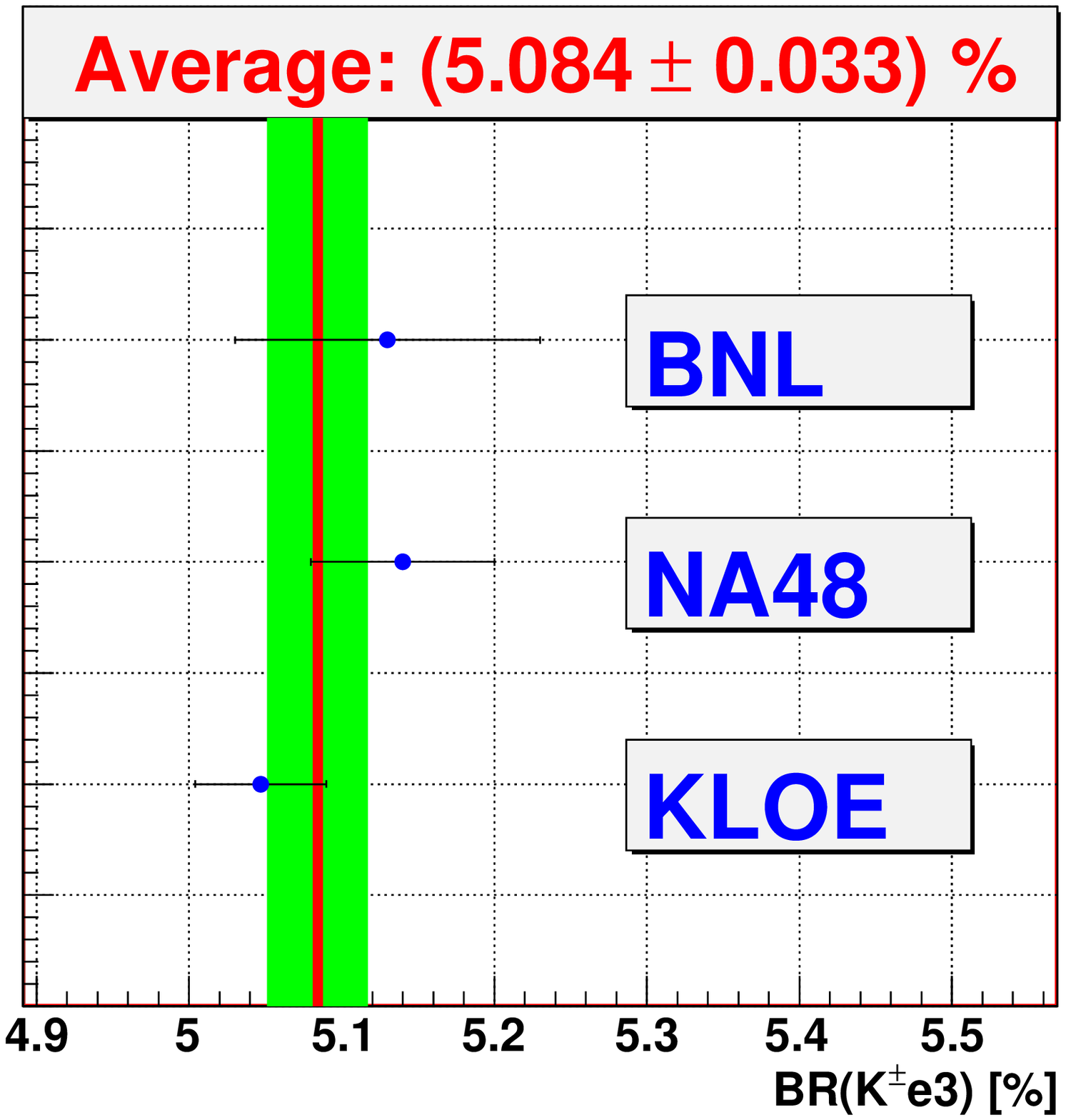}
\includegraphics[width=75mm]{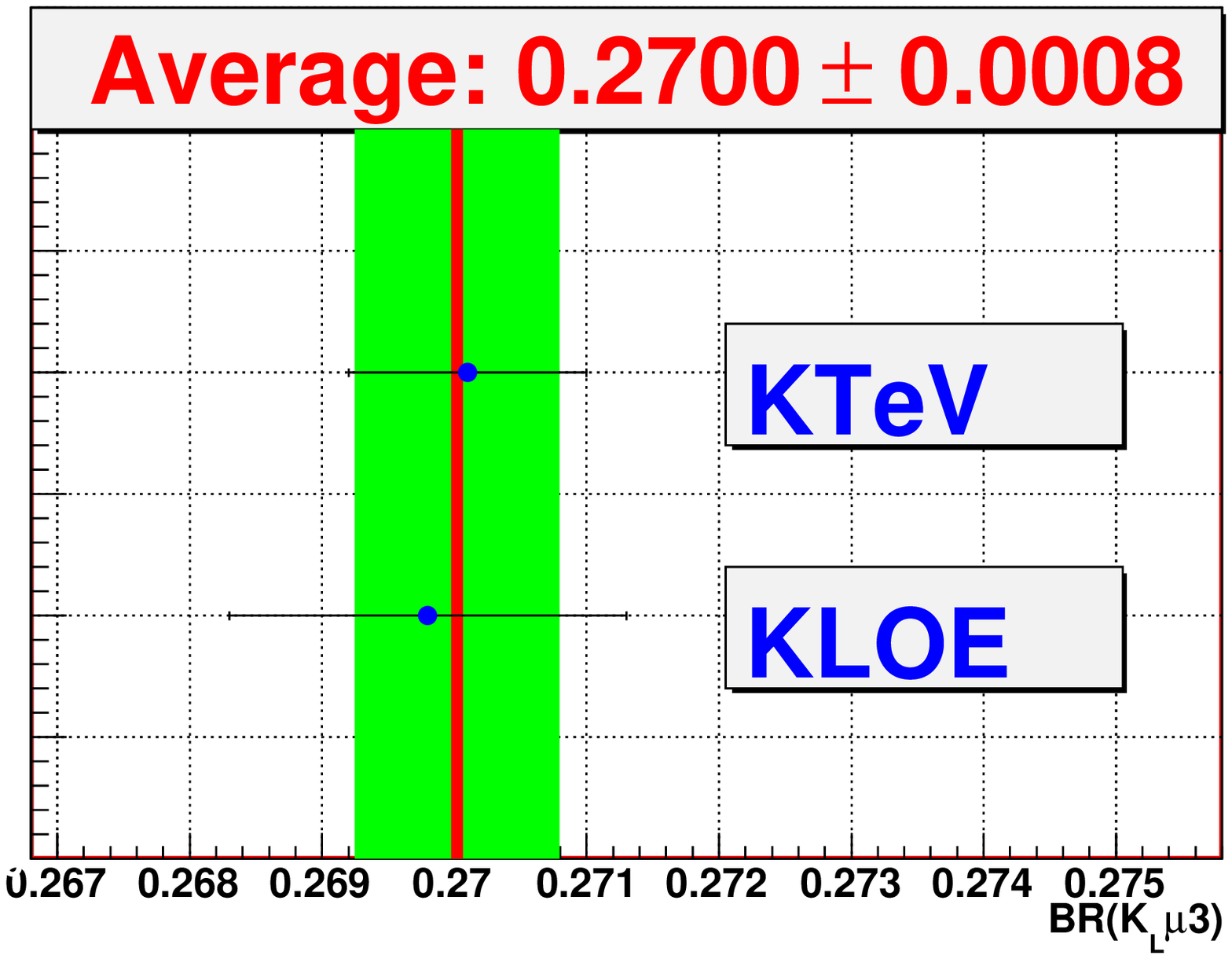}
\includegraphics[width=75mm]{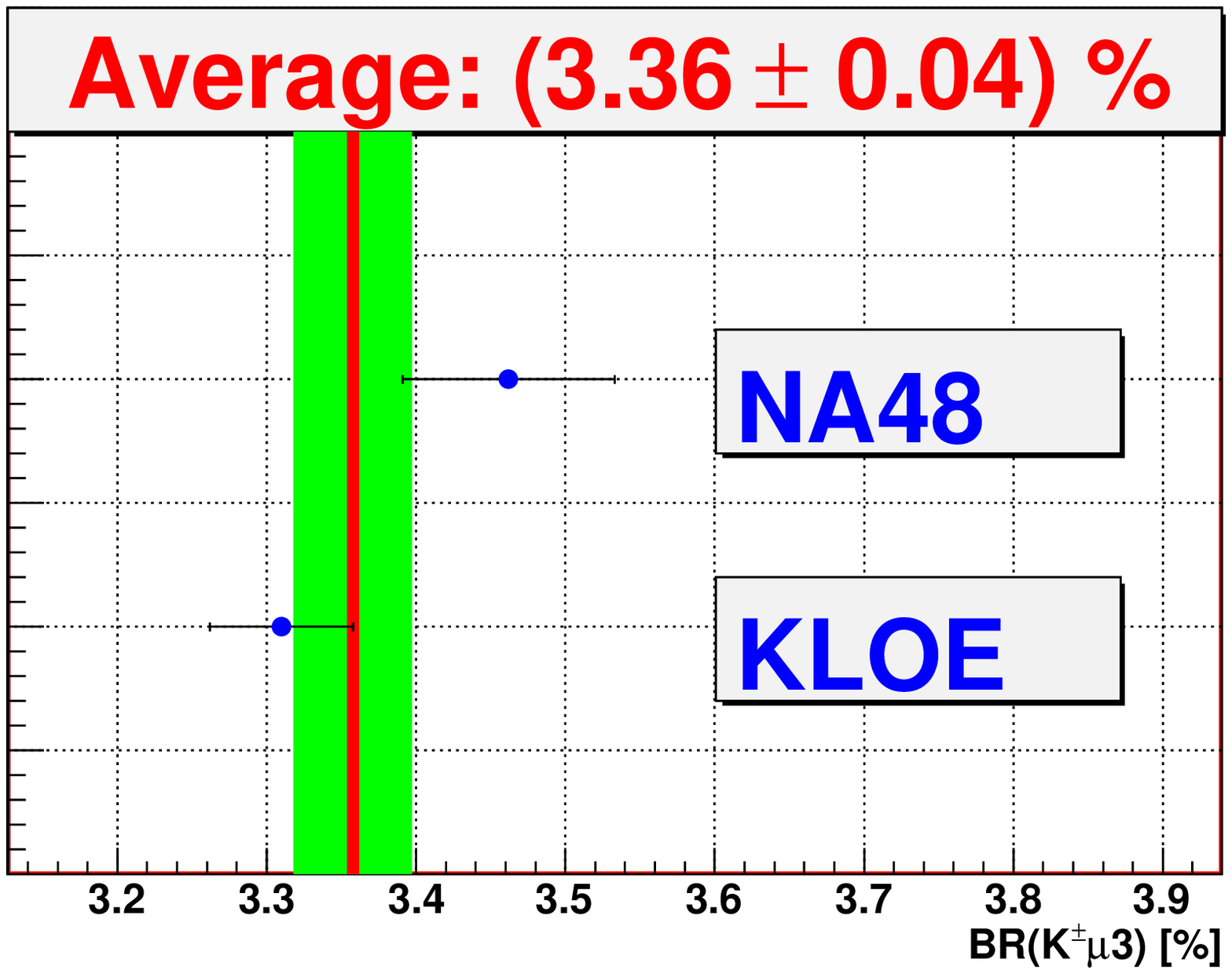}
\caption{Recent measurements of the kaon semileptonic branching ratios. $Br(K_S \to \pi^{\pm} e^{\mp} \nu) = (7.046 \pm 0.091) \%$ }
\label{Kl3br}
\end{figure}

The NA48 experiment has measured the ratio of the branching ratios of $K_L e3$ and all two track events \cite{na48ke3br}. In this way  $Br(K_L e3) = R_e (1.0048 - Br(K_l 3\pi^0)$, where  $Br(K_L 3\pi^0)$ is the external input. Using the measured $R_e = 0.4978 \pm 0.0035 $ and the current PDG value for $Br(K_L 3\pi^0) = (19.69 \pm 0.26) \%$  the result for the $K_L e3$ branching is
\begin{equation}
Br(K_L \to \pi^{\pm} e^{\mp} \nu) = (40.22 \pm 0.31) \% \\
\end{equation}

Preliminary results for the charged semileptonic decays have also been presented by NA48 \cite{na48kche3},\cite{na48kchmu3} and KLOE \cite{kloehql06} 

NA48
\begin{eqnarray}
Br(K^{\pm} \to \pi^0 e^{\pm} \nu )= (5.14 \pm 0.06) \% \\
Br(K^{\pm} \to \pi^0 \mu^{\pm} \nu )= (3.46 \pm 0.07) \%
\end{eqnarray}

KLOE
\begin{eqnarray}
Br(K^{\pm} \to \pi^0 e^{\pm} \nu )= (5.047  \pm 0.043) \% \\
Br(K^{\pm} \to \pi^0 \mu^{\pm} \nu) = (3.310 \pm 0.048) \%
\end{eqnarray}
which confirm the discrepancy with the PDG observed by BNL. 

This ten new measurements of the kaon semileptonic branching ratios together with the BNL result for $Br(K^{\pm}e3) $ are averaged depending on the decay mode and are shown on Figure \ref{Kl3br} (apart from $Br(K_S e3)$, measured only by KLOE). 
As can be seen they show very good consistency. 

\subsection{$\vus$ from kaon semileptonic decays}
\begin{figure}[!ht]
\begin{center}
\includegraphics[width=100mm]{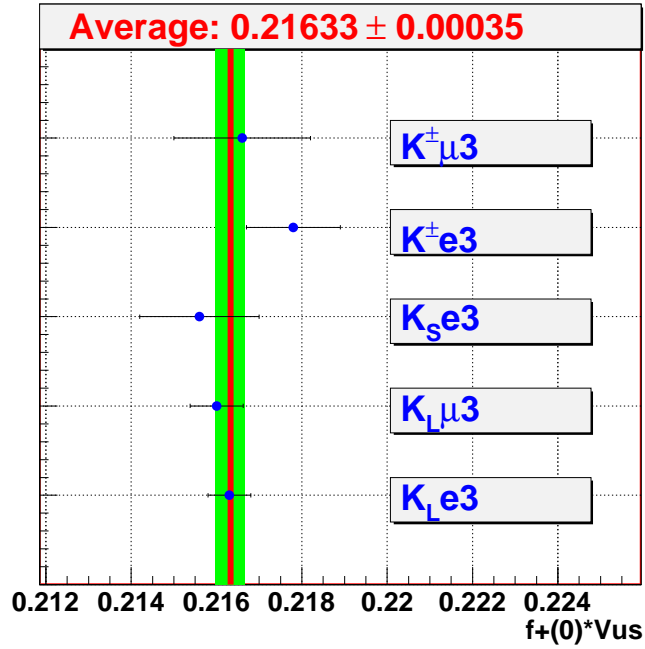}
\end{center}
\caption{The experimentally measured quantity $\vus\times f_+(0)$ from kaon semileptonic decays}
\label{fplusvus}
\end{figure}

Combining all the inputs mentioned above the values for $\vus\times f_+(0)$ from the different modes together with the average are shown of Figure \ref{fplusvus}. 

The precision on the combined measurement of $\vus\times f_+(0)$  is approximately 0.16\%. Using for $f_+(0)$ the value obtained by Leutwyler and Roos the result for $\vus$ is
\begin{equation}
\vus = 0.2251 \pm 0.0019
\label{vus_semil}
\end{equation}
where the dominant contribution to the error comes from the uncertainty of $f_+(0)$. 

\section{$\vus$ from $Kl2$ decays}
A complementary way to extract $\vus$ is to use the ratio of the branching ratios of the pion and the kaon leptonic decays \cite{ckcpi}.  It can be written as
\begin{equation}
 \frac{Br(K^{\pm}\to \mu^{\pm}\nu(\gamma))}{Br(\pi^{\pm}\to \mu^{\pm}\nu(\gamma))}  =  \frac{|\vus|^2}{|\vud|^2} \frac{f_K^2}{f_{\pi}^2} \times \frac{\tau_K}{\tau_{\pi}} \frac{M_K(1-\frac{M_{\mu}^2}{M_K^2})^2}{M_{\pi} (1-\frac{M_{\mu}^2}{M_{\pi}^2})^2} \times \frac{1+\frac{\alpha}{\pi}C_K}{1+\frac{\alpha}{\pi}C_{\pi}}
\end{equation} 
where $\tau_{K,\pi}$ and $f_{K,\pi}$ are the meson lifetimes and decay constants correspondingly and $C_{K,\pi}$ parametrize the electroweak correction.  Using the new measurement of $Br(K^{\pm}\to \mu^{\pm}\nu(\gamma)) = (63.66 \pm 0.17) \% $ from KLOE \cite{kloekmu2} and the lattice QCD calculation of $f_K/f_{\pi}$ \cite{fkfpi} we get $|\vus| / |\vud| = 0.2286^{+0.0026}_{-0.0014}$
which together with the measurement of $\vud$  \cite{vud_marsir} gives
\begin{equation}
\vus = 0.2223^{+0.0026}_{-0.0014} 
\end{equation}
The accuracy of the result is comparable to (\ref{vus_semil}). The dominant error comes from the uncertainty on the ratio  $f_K/f_{\pi}$. 

\section{Conclusions}

The values of $\vus$ extracted from kaon semileptonic decays and  from $K\mu2$ decay agree. The average is 
\begin{equation}
\vus = 0.2241 \pm 0.0015 .
\end{equation}

Using  $\vud = 0.97377(27) $ we have
\begin{equation}
|\vud|^2+|\vus|^2 = 0.9985 \pm 0.0009.
\end{equation}

This result is compatible with the Standard Model and the unitarity of the CKM matrix. 

\section*{Acknowledgements}

I would like to thank prof. dr. Leandar Litov for the valuable help during the preparation of this review and the Joint Institute for Nuclear Research - Dubna for the sponsorship.
\par \bigskip



\begin{thebibliography}{99}


\bibitem{c}
N.~Cabibbo, Phys. Rev. Lett. {\bf 10} (1963) 531.

\bibitem{km}
M. Kobayashi and T.~Maskawa, Prog. Th. Phys. {\bf 49} (1973) 652.

\bibitem{pdg2004}
S. Eidelman, \etal, Phys. Lett. {\bf B 592}  (2004) 1

\bibitem{epove}
  A.~Lai {\it et al.}  [NA48 Collaboration],
  Eur.\ Phys.\ J.\ C {\bf 22} (2001) 231
\\
  A.~Alavi-Harati {\it et al.}  [KTeV Collaboration],
  Phys.\ Rev.\ D {\bf 67} (2003) 012005
  [Erratum-ibid.\ D {\bf 70} (2004) 079904]

\bibitem{ktevtechnical}
http://kpasa.fnal.gov:8080/public/ktev.html

\bibitem{na48technical}
http://na48.web.cern.ch/NA48/

\bibitem{kloetechnical}
http://www.lnf.infn.it/kloe/

\bibitem{marsir}
W. J. Marciano, A. Sirlin, Phys. Rev. Lett. {\bf 71} (1993) 3629

\bibitem{ciriglianokch}
  V.~Cirigliano, M.~Knecht, H.~Neufeld, H.~Rupertsberger and P.~Talavera,
  Eur.\ Phys.\ J.\ C {\bf 23} (2002) 121
  [arXiv:hep-ph/0110153].

\bibitem{ciriglianok0}
  V.~Cirigliano, H.~Neufeld and H.~Pichl,
  Eur.\ Phys.\ J.\ C {\bf 35} (2004) 53

\bibitem{ffs}
H. W. Fearing, E. Fischbach, J. Smith, Phys. Rev. D {\bf 2} (1970) 542.

\bibitem{leutroos}
  H.~Leutwyler and M.~Roos,
  Z.\ Phys.\ C {\bf 25} (1984) 91.


\bibitem{fpluslattice}
  D.~Becirevic {\it et al.},
  Nucl.\ Phys.\ B {\bf 705} (2005) 339

\bibitem{na48ff}
  A.~Lai {\it et al.}  [NA48 Collaboration],
  Phys.\ Lett.\ B {\bf 604} (2004) 1

\bibitem{ktevff}
  T.~Alexopoulos {\it et al.}  [KTeV Collaboration],
  Phys.\ Rev.\ D {\bf 70} (2004) 092007

\bibitem{kloeff}
  F.~Ambrosino {\it et al.}  [KLOE Collaboration],
  Phys.\ Lett.\ B {\bf 636} (2006) 166

\bibitem{ktevps}
  E.~Abouzaid {\it et al.}  [KTeV Collaboration],
  Phys.\ Rev.\ D {\bf 74} (2006) 097101

\bibitem{istraff}
  O.~P.~Yushchenko {\it et al.},
  Phys.\ Lett.\ B {\bf 589} (2004) 111


\bibitem{kloelife}
  F.~Ambrosino {\it et al.}  [KLOE Collaboration],
  Phys.\ Lett.\ B {\bf 626} (2005) 15

\bibitem{kloeklbr}
  F.~Ambrosino {\it et al.}  [KLOE Collaboration],
  Phys.\ Lett.\ B {\bf 632} (2006) 43

\bibitem{kloehql06}
  R.~Versaci  [By KLOE Collaboration],
  arXiv:hep-ex/0701008.

\bibitem{pdg06}
  W.-M. Yao, \etal Journal of Physics G {\bf 33} (2006) 1

\bibitem{bnlke3}
  A.~Sher {\it et al.},
  Phys.\ Rev.\ Lett.\  {\bf 91} (2003) 261802

\bibitem{pdg02}
  K. Hagiwara \etal, Physical Review {\bf D 66} (2002) 010001

\bibitem{ktevbrall}
  T.~Alexopoulos {\it et al.}  [KTeV Collaboration],
  Phys.\ Rev.\ D {\bf 70} (2004) 092006

\bibitem{kloeks}
  F.~Ambrosino {\it et al.}  [KLOE Collaboration],
  Phys.\ Lett.\ B {\bf 636} (2006) 173

\bibitem{na48ke3br}
  A.~Lai {\it et al.}  [NA48 Collaboration],
  Phys.\ Lett.\ B {\bf 602} (2004) 41

\bibitem{na48kche3}
  L.~Litov  [NA48 Collaboration],
  arXiv:hep-ex/0501048.

\bibitem{na48kchmu3}
  A. Dabrowski, presented at KAON 2005 Workshop

\bibitem{ckcpi}
  W.~J.~Marciano,
  Phys.\ Rev.\ Lett.\  {\bf 93} (2004) 231803

\bibitem{kloekmu2}
  F.~Ambrosino {\it et al.}  [KLOE Collaboration],
  Phys.\ Lett.\ B {\bf 632} (2006) 76

\bibitem{fkfpi}
  C.~Bernard {\it et al.}  [MILC Collaboration],
  arXiv:hep-lat/0609053.

\bibitem{vud_marsir}
  W.~J.~Marciano and A.~Sirlin,
  Phys.\ Rev.\ Lett.\  {\bf 96} (2006) 032002

\end{thebibliography}
\end{document}